\documentclass[letterpaper, 10 pt, conference]{ieeeconf}  

\IEEEoverridecommandlockouts                              
\usepackage{graphics} 
\usepackage{color}
\usepackage{tikz}
\usepackage{subfig}
\usepackage{graphicx}
\usepackage{xcolor}
\usepackage{amsmath} 
\usepackage{amssymb}  
\usepackage{placeins}
\usepackage[normalem]{ulem}
\usepackage{caption}
\usepackage{titlesec}
\definecolor{blue_def}{HTML}{1f77b4}
\definecolor{orange_def}{HTML}{ff7f0e}
\definecolor{green_def}{HTML}{2ca02c}
\definecolor{red_def}{HTML}{d62728}
\newcommand{\redline}{\raisebox{-1pt}{\tikz{\draw[-,red_def,solid,line width = 6](0,0) -- (3.5mm,0);}}}
\newcommand{\orangeline}{\raisebox{-1pt}{\tikz{\draw[-,orange_def,solid,line width = 6](0,0) -- (3.5mm,0);}}}
\newcommand{\greenline}{\raisebox{-1pt}{\tikz{\draw[-,green_def,solid,line width = 6](0,0) -- (3.5mm,0);}}}
\newcommand{\blueline}{\raisebox{-1pt}{\tikz{\draw[-,blue_def,solid,line width = 6](0,0) -- (3.5mm,0);}}}

\titlespacing{\section}{3pt}{\parskip}{-\parskip}
\titlespacing{\subsection}{3pt}{\parskip}{-\parskip}
\titlespacing{\subsubsection}{3pt}{\parskip}{-\parskip}

\title{\LARGE \bf
Deep Identification of Nonlinear Systems in Koopman Form
}

\author{Lucian Cristian Iacob, Gerben Izaak Beintema, Maarten Schoukens and Roland Tóth
\thanks{This work has received funding from the European Research Council (ERC) under the European Union’s Horizon 2020 research and innovation programme (grant agreement nr. 714663).}
\thanks{The authors are with the Control System Group,
        Eindhoven University of Tehcnology, The Netherlands
        {\tt\small \{l.c.iacob, g.i.beintema, r.toth, m.schoukens\}}\tt\small@tue.nl}
        \thanks{Roland Tóth is also with the Systems and Control Laboratory, Institute for Computer Science and Control, Budapest, Hungary}
}

\begin{document}

\maketitle
\thispagestyle{empty}
\pagestyle{empty}

\begin{abstract}
The present paper treats the identification of nonlinear dynamical systems using Koopman-based deep state-space encoders. Through this method, the usual drawback of needing to choose a dictionary of lifting functions a priori is circumvented. The encoder represents the lifting function to the space where the dynamics are linearly propagated using the Koopman operator. An input-affine formulation is considered for the lifted model structure and we address both full and partial state availability. The approach is implemented using the the deepSI toolbox in Python. To lower the computational need of the simulation error-based training, the data is split into subsections where multi-step prediction errors are calculated independently. This formulation allows for efficient batch optimization of the network parameters and, at the same time, excellent long term prediction capabilities of the obtained models. The performance of the approach is illustrated by nonlinear benchmark examples.
\end{abstract}

\smallskip
\section{INTRODUCTION}
\smallskip
 Nonlinear system identification is a wide and intensely researched topic, aiming at the estimation of dynamical systems directly form data. Multiple methods have been developed, where the commonly used model structures are: NARX, nonlinear state space, block-oriented, see \cite{Schoukens_NI} for an overview. However, even if the resulting models have good simulation or prediction performance, the representation of the system remains confined in the nonlinear system class. While several nonlinear control methods have been developed (e.g. feedback linearization, backstepping, sliding mode control, to name a few \cite{khalil_nl}), they are often complicated to apply and there is no systematic approach for shaping the performance of the closed-loop system in contrast to the approaches of the linear time invariant (LTI) framework. While LTI control is well advanced, designs are limited to operate in the neighbourhood of given linearization points. Hence, pressing needs in engineering led to the idea of developing various linear embeddings of nonlinear systems to apply the powerful LTI control methods with global stability and performance guarantees.
 \par One such embedding technique is based on the Koopman framework, where the concept is to lift the nonlinear state space to a (possibly) infinite dimensional space through so-called observable functions. The dynamics of the original system are preserved and governed by the linear Koopman operator \cite{book_koopman}, \cite{brunton_overview}. In practice, if a dictionary of a finite number of observables is chosen a priori and used to construct time shifted data matrices, the linear Koopman-based model can be calculated via least squares \cite{Mauroy_sysid}. One such approach, called Dynamic Mode Decomposition (DMD) \cite{rowley_dmd}, is based on constructing the time shifted data matrices using the original states of the system. If the dictionary consists of nonlinear functions of the state, this technique is known as Extended DMD (EDMD) \cite{williams_edmd}.   Besides issues that arise with the presence of noise and biasedness of the estimates, the main difficulty lays with choosing the lifting functions such that, on the lifted state-space, an LTI model exists that can well capture the dynamic behavior of the original nonlinear system.
\par Learning the lifting functions from data has been addressed recently using Artificial Neural Networks (ANNs). A common approach is to construct autoencoders, to represent the lifting function and its inverse, \cite{lusch_nn}-\cite{Otto}, and enforce the linearity condition of the  lifted state transition. Alternatively, \cite{yeung_nn} proposes to specify the entire dictionary of observables as the outputs of an ANN, and perform an EDMD type of regression for model estimation. While \cite{yeung_nn} addresses partial state observations, availability of full state measurements is a common assumption in the Koopman identification literature. Moreover, only a few papers present examples where measurement noise is present (e.g. \cite{takeishi_noise}) and often only the robustness of the methods is analysed instead of ensuring stochastic consistency of the estimators. Furthermore, in this research area, the treatment of inputs has only recently been addressed, either through a nonlinear lift \cite{bonnert_nn} or by using state and input dependent observables, together with input increments \cite{deepkoco_nn}.
\par To address these issues, we introduce a Koopman-based state-space encoder model and a corresponding estimation method, implemented based on the deepSI toolbox\footnote{deepSI toolbox available at https://github.com/GerbenBeintema/deepSI} in Python \cite{Gerben_l4dc}. We summarize the main characteristics and contributions as follows:
\begin{itemize}
\item Constructability-based formulation of a lifted space encoder (nonlinear mapping) based on deep-ANN
\item Formulation of an identification approach for estimating Koopman models with Output Error (OE) noise structure in state-space using $T$-step ahead prediction
\item Incorporation of lifted-state dependent linear effect of the input for general representation of nonlinear systems
\item Construction that allows for both full and partial state availability
\item Computationally scalable implementation of the estimation via batch-wise (multiple-shooting) optimization of a single and relatively simple loss function, compared to the more complex training criteria in \cite{lusch_nn}, \cite{deepkoco_nn}.
\end{itemize}
\par The paper is structured as follows. Section \ref{section_behavior} details the general Koopman framework and we discuss the notions of observability and state constructability in the Koopman form together with the role of inputs. Section \ref{section_encoder} describes the proposed Koopman-based encoder and the associated model structure together with the used optimization method. In Section \ref{section_experiments}, the approach is tested using a Van der Pol oscillator and the Silverbox benchmark \cite{Silverbox}, followed by a discussion of the results. The conclusions are presented in Section \ref{section_conclusion}.
\vspace{.1cm}
\section{Behavior of Koopman embeddings}\label{section_behavior}
\vspace{.1cm}
This section details the Koopman framework focusing on a finite dimensional lifted form.  Next, we briefly discuss observability and constructability notions in the original and lifted forms. We show that, while the system behavior can be represented by a linear form, a nonlinear constraint remains on the initial conditions to ensure one-to-one connection between the solution sets.  \vspace{.1cm}
\subsection{Preliminaries}
\vspace{.1cm}
Consider a discrete-time nonlinear autonomous system:
\begin{equation}\label{eq:nl_aut}
x_{k+1}=f(x_k)
\end{equation}
with $x:\mathbb{Z}\rightarrow\mathbb{R}^{n_\mathrm{x}}$ the state variable, $f:\mathbb{R}^{n_\mathrm{x}}\rightarrow\mathbb{R}^{n_\mathrm{x}}$ is a bounded nonlinear function and $k\in\mathbb{Z}$ is the discrete time step. The Koopman framework proposes an alternative representation of system \eqref{eq:nl_aut} by introducing so-called observable functions $\phi\in\mathcal{F}$, with $\mathcal{F}$ a Banach function space and $\phi:\mathbb{R}^{n_\mathrm{x}}\rightarrow \mathbb{R}$. As described in \cite{book_koopman}, the Koopman operator $\mathcal{K}:\mathcal{F}\rightarrow\mathcal{F}$ associated with \eqref{eq:nl_aut} and $\mathcal{F}$ is defined through:
\begin{equation}\label{eq:koop_composition}
\mathcal{K}\phi=\phi\circ f, \quad \forall \phi \in \mathcal{F}
\end{equation}
where $\circ$ denotes function composition and \eqref{eq:koop_composition} is equal to:
\begin{equation}
    \mathcal{K}\phi(x_k)=\phi(x_{k+1}).
\end{equation}
An important property of the Koopman operator is that it is linear when $\mathcal{F}$ is a vector space of functions \cite{brunton_overview}. Considering two observables $\phi_1,\phi_2\in\mathcal{F}$ and scalars $\alpha_1,\alpha_2\in\mathbb{R}$, if \eqref{eq:koop_composition} holds, then it implies: \vspace{.1cm}
\begin{equation}
\begin{split}
    \mathcal{K}(\alpha_1\phi_1 + \alpha_2\phi_2)&=(\alpha_1\phi_1+\alpha_2\phi_2)\circ f\\
    &=\alpha_1\phi_1\circ f + \alpha_2\phi_2 \circ f\\
    &=\alpha_1\mathcal{K}\phi_1 + \alpha_2\mathcal{K}\phi_2,
    \end{split}\vspace{.2cm}
\end{equation}
proving the linearity property. While often the existence of the Koopman operator requires $\mathcal{F}$ to be spanned by an infinite number of basis functions, for practical purposes, an $n_\mathrm{f}$-dimensional linear subspace $\mathcal{F}_{n_\mathrm{f}}\subset \mathcal{F}$ is considered, with $\mathcal{F}_{n_\mathrm{f}}=\mathrm{span}\left\lbrace\phi_j\right\rbrace^{n_\mathrm{f}}_{j=1}$. As detailed in \cite{book_koopman}, with a projection operator $\Pi:\mathcal{F}\rightarrow\mathcal{F}_{n_\mathrm{f}}$, the finite-dimensional approximation of the Koopman operator $\mathcal{K}$ is given by:\vspace{.1cm}
\begin{equation}
\mathcal{K}_{n_\mathrm{f}}=\left.\Pi \mathcal{K}\right|_{\mathcal{F}_{n_\mathrm{f}}}: \mathcal{F}_{n_\mathrm{f}} \rightarrow \mathcal{F}_{n_\mathrm{f}}.\vspace{.1cm}
\end{equation}
In practice, the Koopman matrix representation $A\in\mathbb{R}^{n_\mathrm{f}\times n_\mathrm{f}}$ 
\vfill\break
\noindent{is used, such that the element-wise relation is satisfied:}
\begin{equation}
\mathcal{K}_{n_\mathrm{f}} \phi_{j} = \sum_{i=1}^{n_\mathrm{f}} A_{ji} \phi_{i}.
\end{equation}
For a more detailed analysis, one can consult \cite{book_koopman}. Next, we introduce the lifted state $z_k = \Phi(x_k)$, where $\Phi(x_k)=\begin{bmatrix}
\phi_1(x_k) & \dots &  \phi_{n_\mathrm{f}}(x_k)
\end{bmatrix}^\top$. The lifted finite dimensional linear representation of \eqref{eq:nl_aut} is then given by:\vspace{.1cm}
\begin{equation}\label{eq:koop_aut}
z_{k+1}=Az_k. \vspace{.1cm}
\end{equation}
However, the main difficulty of the Koopman framework is the choice of the lifting functions such that the resulting observables generate a Koopman invariant subspace \cite{Brunton_2016}. Furthermore, it is often not clearly stated in the literature on the subject that  a linear system whose dynamics are driven by the Koopman matrix $A$ is only equivalent in terms of behavior (collections of all solution trajectories) with the original nonlinear system \eqref{eq:nl_aut} if explicit nonlinear constraints are defined for the initial condition of the lifted state. We explore this next using a simple example. \vspace{.1cm}
\subsection{Linear representations subject to nonlinear constraints}\label{sec_2b}
\vspace{.1cm}
To illustrate the concept, consider a nonlinear system represented by \eqref{eq:nl_aut} with a polynomial $f$, similar to the one described in \cite{Brunton_2016}. 
Denote $x_{k,i}$ to be the $i^{\text{th}}$ element of $x_k$. In this notation, the system dynamics are described as follows:\vspace{.1cm}
\begin{equation}\label{eq:example_aut}
\begin{bmatrix}
x_{k+1,1} \\x_{k+1,2}
\end{bmatrix}=\begin{bmatrix}
ax_{k,1} \\ bx_{k,2} - cx^2_{k,1}
\end{bmatrix}\vspace{.1cm}
\end{equation}
with constant parameters $a,b,c\in\mathbb{R}$. By considering solutions of \eqref{eq:example_aut} only on $[0,\infty )$ with initial condition $x_0\in\mathbb{R}^{n_2}$, the feasible trajectories are given by:\vspace{.1cm}
\begin{equation}
\mathcal{B} =\left\lbrace x:\mathbb{Z}^+_0 \rightarrow\mathbb{R}^2\mid \text{s.t. \eqref{eq:example_aut} is satisfied}\right\rbrace.
\vspace{.1cm}
\end{equation}
To represent the system in the Koopman form, the following observables are chosen: $\phi_1(x_k) = x_{k,1}$, $\phi_2(x_k) = x_{k,2}$ and $\phi_3(x_k) = x^2_{k,1}$. Then, the dynamics are represented by:  \vspace{.05cm}
\begin{equation}\label{eq:lifted_phi_ex}
\Phi(x_{k+1})=\underbrace{\begin{bmatrix}
a & 0 & 0 \\ 0 & b & -c \\ 0 & 0 & a^2
\end{bmatrix}}_A \Phi(x_k). \vspace{.05cm}
\end{equation} 
Based on \eqref{eq:lifted_phi_ex}, consider the system $z_{k+1}=Az_k$ of dimension $n_\mathrm{z}=3$, with $z_0\in\mathbb{R}^3$; the solution set is described as: 
\begin{equation}\label{eq:lifted_z_set}
\mathcal{B}_\mathcal{K}=\left\lbrace z : \mathbb{Z}^+_0\rightarrow\mathbb{R}^3\mid \text{s.t. }z_{k+1}=Az_k\right\rbrace.
\end{equation}
Note that \eqref{eq:lifted_z_set} represents an unrestricted LTI behavior. It is easy to show that $\Phi(\mathcal{B})\subseteq\mathcal{B}_\mathcal{K}$, as any $z_k\in\mathcal{B}_{\mathcal{K}}$ with $z_0\in\mathbb{R}^3$ for which $z_{0,3}\neq z^2_{0,1}$ will not correspond to a solution of \eqref{eq:example_aut}, i.e. $\Phi^{-1}(z_k)=x_k\notin\mathcal{B}$. By introducing the constraint $\Psi:\mathbb{R}^3\rightarrow\mathbb{R},\Psi(z_k)=z^2_{k,1}-z_{k,3}$, the solution set \eqref{eq:lifted_z_set} with constraint $\Psi$ is: \vspace{.1cm}
\begin{equation}
\hat{\mathcal{B}}_{\mathcal{K}}=\left\lbrace z : \mathbb{Z}^+_0\rightarrow\mathbb{R}^3\mid\text{s.t. }z_{k+1}=Az_k,\;\Psi(z_0)=0\right\rbrace .
\vspace{.1cm}
\end{equation}
Then, it is possible to show that $\Phi(\mathcal{B})=\hat{\mathcal{B}}_\mathcal{K}$. Our example shows that, to have a bijective relation between the solution sets, additional constraints need to be imposed on the Koopman form, or, as we call it now, embedding of \eqref{eq:nl_aut}.
\subsection{Observability, constructability and extension to inputs}
\vspace{.1cm}
\subsubsection{Autonomous case}
\vspace{.1cm}
Consider the system \eqref{eq:nl_aut} having the output defined as:
\begin{equation}\label{eq:nl_out}
    y_k = h(x_k), \vspace{-.1cm}
\end{equation}
with the nonlinear output map $h:\mathbb{R}^{n_\mathrm{x}}\rightarrow\mathbb{R}^{n_\mathrm{y}}$. Given $x_0\in\mathbb{R}^{n_\mathrm{x}}$, the observability map for the nonlinear system represented by \eqref{eq:nl_aut} and \eqref{eq:nl_out} is: 
\begin{equation}
\mathcal{O}_\mathrm{x}(x_0)=\begin{bmatrix}
h(x_0) \\ h^{(1)}(x_0) \\ \dots \\ h^{(n_\mathrm{x}-1)}(x_0)
\end{bmatrix}=\begin{bmatrix}
y_0 \\ y_{1} \\ \dots \\ y_{n_\mathrm{x}-1}
\end{bmatrix}
\end{equation}
with $h^{(i)}(x_0)=h(f^{(i)}(x_0))$ and $f^{(i)}$ is the composition of $f$ with itself $i$ times. As described in \cite{nl_obs_1982}, the system satisfies the observability rank condition at $x_0$ if the rank of the analytical map\footnote{The rank of the Jacobian matrix of $\mathcal{O}_\mathrm{x}$ at $x_0$.} $\mathcal{O}_\mathrm{x}:\mathbb{R}^{n_\mathrm{x}}\rightarrow\mathbb{R}^{n_\mathrm{x} n_\mathrm{y}}$ is equal to $n_\mathrm{x}$. If this condition is met, the system is strongly locally observable at $\bar{x}\in X$, where $X$ is a neighbourhood of $x_0$ and there exists an analytic function $\mathcal{O}^{-1}_\mathrm{x}:\mathbb{R}^{n_\mathrm{x} n_\mathrm{y}}\rightarrow X$, such that $\mathcal{O}^{-1}_\mathrm{x}\left(\begin{bmatrix}
y^\top_0 & \dots & y^\top_{n_\mathrm{x}-1}
\end{bmatrix}^\top\right)=x_0$.  \vspace{.01cm}
\par In the Koopman form, assuming that the output function is in the span of the lifted states, i.e., $y_k=Cz_k$, with $C\in\mathbb{R}^{n_\mathrm{y}\times n_\mathrm{z}}$, the observability map can be defined as follows: 
\begin{equation}\label{eq:nl_alg_constr}
\begin{bmatrix}
y_0 \\ y_{1} \\ \dots \\ y_{n_\mathrm{z}-1} \\ 0
\end{bmatrix} =\begin{bmatrix}\begin{pmatrix}
C \\ CA \\ \dots \\ CA^{n_\mathrm{z}-1} \end{pmatrix}z_0 \\ \Psi(z_0) 
\end{bmatrix}=\mathcal{O}_\mathrm{z}(z_0)
\end{equation}
where, as observed in section \ref{sec_2b}, it is also necessary to consider the nonlinear constraints $\Psi:\mathbb{R}^{n_\mathrm{f}}\rightarrow\mathbb{R}^{n_\mathrm{c}}$. The map $\mathcal{O}_{\mathrm{z}}$ should be locally invertible to uniquely determine $z_0$, i.e. there exists $\mathcal{O}^{-1}_{\mathrm{z}}:\mathbb{R}^{n_\mathrm{z} n_\mathrm{y}}\rightarrow\mathbb{R}^{n_\mathrm{f}}$, such that $\mathcal{O}^{-1}_z\left(\begin{bmatrix}
y^\top_0 & \dots & y^\top_{n_\mathrm{x}-1}\end{bmatrix}^\top\right)=z_0$. This is a different point of view than in the work \cite{surana_obs}, where the observability notions are discussed based on an explicit definition of the lifting map. Similar to \eqref{eq:nl_alg_constr}, the notion of constructability refers to uniquely determining $z_0$ using current and past measurements, that is, $\mathcal{R}^{-1}_\mathrm{z}\left(\begin{bmatrix}
y^\top_{-n_\mathrm{z}+1} & \dots & y^\top_0 \end{bmatrix}^\top\right)=z_0$ with $\mathcal{R}^{-1}_{\mathrm{z}}:\mathbb{R}^{n_\mathrm{z}  n_\mathrm{y}}\rightarrow\mathbb{R}^{n_\mathrm{f}}$. $\mathcal{R}_\mathrm{z}$ is the constructability map and $\mathcal{R}^{-1}_\mathrm{z}$ denotes its inverse. In the proposed ANN implementation, we formulate the encoder as a nonlinear function in order to reconstruct a state that can be associated with the Koopman form of the nonlinear system. 
\subsubsection{Systems with input} 
\vspace{.1cm}
To extend the considerations to the non-autonomous case, we consider the class of nonlinear control-affine systems:
\begin{equation}\label{eq:nl_au_in_ca}
    x_{k+1}=f(x_k)+g(x_k)u_k,
\end{equation}
with a potential nonlinear function $g:\mathbb{R}^{n_\mathrm{x}}\rightarrow\mathbb{R}^{\mathrm{n_x}\times n_\mathrm{u}}$ and input $u:\mathbb{Z}\rightarrow\mathbb{R}^{n_\mathrm{u}}$. The treatment of the inputs in the lifted form is a topic of debate with many different approaches present in the literature. In general, an LTI form is assumed due to its ease of use with existing linear control methods \cite{korda_mpc}. However, this form may be insufficient to capture the nonlinear behavior of \eqref{eq:nl_au_in_ca}. Based on the results of \cite{surana_obs} for continuous time, we consider an input-affine Koopman form:
\begin{equation}\label{eq:lifted_koop_ia}
    \Phi(x_{k+1}) = A\Phi(x_k) + B(\Phi(x_k))u_k.
\end{equation}
We argue that, although \eqref{eq:lifted_koop_ia} is more complex than an LTI model, it provides a better approximation capability of the dynamics of \eqref{eq:nl_au_in_ca}. 
\par Let $z_k=\Phi(x_k)$ and $y_k=Cz_k$. The observability map $\Theta_z:\mathbb{R}^{n_\mathrm{z} n_\mathrm{u}  n_\mathrm{z}}\rightarrow\mathbb{R}^{n_\mathrm{z} n_\mathrm{y}+n_\mathrm{c}}$ is described as:
\begin{equation*}
\begin{split}
&\left(\Theta_z(z_0,\textbf{u})\right)^\top = \begin{bmatrix}
y^\top_0 & y^\top_1 & y^\top_2 & \dots & y^\top_{n_\mathrm{z}-1} & 0^\top
\end{bmatrix}^\top
\\&=\mathcal{O}_\mathrm{z}(z_0)+\begin{bmatrix}
0 \\ CB(z_0)u_0 \\ CAB(z_0)u_0 + CB(\zeta^0(z_0,u_0))u_1 \\ \dots \\ \zeta^{n_{\mathrm{z}}-1}(z_0, \textbf{u}) \\0
\end{bmatrix}
\end{split}
\end{equation*}
where, for ease of readability, $\zeta^0(z_0,u_0)=Az_0+B(z_0)u_0$, $\zeta^{n_\mathrm{z}-1}(z_0,\textbf{u})$ represents a nonlinear function containing all the remaining terms in the expansion of $y_{n_\mathrm{z}-1}$ and $\textbf{u}=[u_0,\;\dots ,\; u_{n_\mathrm{z}-1}]$. As can be seen, for the lifted form \eqref{eq:lifted_koop_ia}, to determine $z_0$ for future input and output data requires the inversion of an even more complex nonlinear map. The same holds for constructability, where the aim is to determine $z_0$ based on past input-output data. Similar to the autonomous case, the encoder is formulated as a nonlinear function that estimates the inverse of the constructability map to determine the lifted state using past measurements.
\section{Identification framework}\label{section_encoder}
\vspace{.1cm}
Based on the the constructability map and the state-dependent affine mapping for the input discussed in Section \ref{section_behavior}, we have now all the ingredients to develop an identification method for a data-driven Koopman embedding of a nonlinear system without prior selection of the observables. Due to the shown nonlinearity of the constructability map, a deep-ANN-based function estimator is needed to determine the state basis $z$.
\subsection{Data generating system}
\vspace{.1cm}
Similar to \eqref{eq:nl_au_in_ca}, the data generating system is considered to be a nonlinear control affine system:
\begin{subequations}\label{eq:data_gen}
\begin{align}
    x_{k+1}&=f(x_k)+g(x_k)u_k \label{eq:data_gen_x}\\
    y_k &=h(x_k)+v_k \label{eq:data_gen_y}
\end{align}
\end{subequations}
with $v_k\in\mathbb{R}^{n_\mathrm{y}}$ being an additive zero-mean (possibly coloured) noise. The stochastic system in \eqref{eq:data_gen} corresponds to an OE noise setting where our objective is to estimate a model of the deterministic (process) part. Next, we define the chosen model structure for the lifted Koopman form.
\subsection{Model structure}
\vspace{.1cm}
We apply the estimation concept from \cite{Gerben_l4dc}, \cite{Gerben_paper} and develop an implementation of the resulting method using deepSI. To represent the input contribution in the lifted model, we consider an input-affine formulation. The chosen model structure is: \vspace{-.1cm}
\begin{subequations}\label{eq:model_structure}
\begin{align}
    \hat{z}_{k+1}&=A_\theta\hat{z}_k+B_\theta(\hat{z}_k)u_k\label{eq:model_structure_1}\\
    \hat{y}_k&=C_\theta\hat{z}_k\label{eq:model_structure_2}
\end{align}
\end{subequations}
where $\hat{z}_k\in\mathbb{R}^{n_\mathrm{z}}$ is the lifted state, $u_k\in\mathbb{R}^{n_\mathrm{u}}$ is the input, $\hat{y}_k\in\mathbb{R}^{n_\mathrm{y}}$ is the model output and $y_k$ is the measured system output. In the proposed model \eqref{eq:model_structure}, $A_\theta$ is the Koopman matrix and $B_\theta (z)$ is a nonlinear function of the lifted state. We construct the model using a linear output map $C_\theta$ such that the outputs of the original model are spanned by the lifting functions. If state measurements are available, we can also enforce that the states can be recovered by a linear mapping. The neural network is constructed using the encoder function $e_\theta$ and the nonlinear map $B_\theta$ (both implemented using feedforward neural nets), and the linear maps $A_\theta$ and $C_\theta$.  The subscript $\theta$ represents the parameters (weights and biases) of the neural network. The main advantage of using the Koopman model structure \eqref{eq:model_structure_1} is that it can be viewed as a Linear Parameter Varying (LPV) system, for which numerous control techniques have been developed (see \cite{lpv_book}).
\par The orders $n_a$ and $n_b$ and their selection corresponds to the classical problem of model structure selection in system identification and hence it is out of the scope of the current paper. Furthermore, the proposed network architecture can also handle full state measurements, in this case the output function $h$ being an identity function.
\begin{figure}
  \centering
\vbox{\vspace{+1em}\includegraphics[scale=0.525]{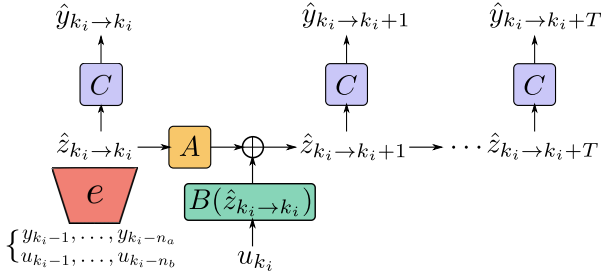}} 
  \caption{Network architecture. The lifted state at moment $k_i$, $\hat{z}_{k_i\rightarrow k_i}$, is estimated using the encoder function $e$ based on previously measured input and output data.\vspace{-.4cm}} \label{fig:encoder_structure}
\end{figure}
\subsection{Cost function and estimation procedure}
\vspace{.1cm}
The computation of the simulated response and the corresponding gradient for ANN optimization has a heavy computational cost, which can become intractable when large data sets are used. To deal with this, a trade-off proposed in \cite{Gerben_l4dc}, \cite{Gerben_paper} is to construct a cost function using subsections of the data set and, starting from an index $k_i$, perform a $T$-step ahead prediction. The cost function is formulated as follows: 
\begin{subequations}
\begin{align}
 V_{\text {enc}}(\theta) =&\frac{1}{2 N (T+1)} \sum_{i=1}^{N} \sum_{p=0}^{T}\left\|\hat{y}_{k_{i} \rightarrow k_{i}+p}-y_{k_{i}+p}\right\|_{2}^{2} \label{eq:cost_enc}\\ 
 \hat{y}_{k_{i} \rightarrow k_{i}+p} &:=C_\theta\hat{z}_{k_{i} \rightarrow k_{i}+p} \\ 
\hat{z}_{k_{i} \rightarrow k_{i}+p+1} &:=A_\theta\hat{z}_{k_i\rightarrow k_i +p} + B_\theta (\hat{z}_{k_i\rightarrow k_i +p})u_{k_i + p}
\end{align}
 \end{subequations}
where $\hat{z}_{k_i\rightarrow k_i+p}$ is computed through $p$ recursive iterations of \eqref{eq:model_structure_1} starting from $\hat{z}_{k_i\rightarrow k_i}$. The initial lifting to $\hat{z}_{k_i\rightarrow k_i}$ is performed through the encoder function $e_\theta$ as follows:
\begin{subequations}
\begin{align}
     \hat{z}_{k_{i} \rightarrow k_{i}} &:=e_\theta\left(x_{k_i-1},u_{k_i-1}\right)\label{eq:zk_fullstate} \\
 \hat{z}_{k_{i} \rightarrow k_{i}} &:=e_\theta(y_{k_i-n_a:k_i-1},u_{k_i-n_b:k_i-1}),\label{eq:zk_io}
\end{align}
\end{subequations}
where $e_\theta$ estimates the inverse of the constructability map, using past input-output data to determine the lifted state $\hat{z}_{k_i\rightarrow k_i}$. The notation $y_{k_i-n_a:k_i-1},u_{k_i-n_b:k_i-1}$ represents the sets of past outputs and inputs. In the case of full state availability, for numerical reasons, the initial lifted state $\hat{z}_{k_i\rightarrow k_i}$ is computed via the encoder function $e_\theta$ based on the previous time step of the original state $x_{k_i-1}$ and input $u_{k_i-1}$ \eqref{eq:zk_fullstate}. 
\subsection{Batch optimization}
\vspace{.1cm}
As detailed in \cite{Gerben_l4dc}, eq. \eqref{eq:cost_enc} allows for the parallelization of the computations, allowing for the cost of each section to be computed individually. As such, the computation time is greatly decreased and, moreover, a batch cost function can be utilized, summing over only a subset of sections:
\begin{align} 
\begin{split}
V_{\text {batch }}(\theta) &=\frac{1}{2 N_{\text {batch }}(T+1)} \sum_{i \in B} \sum_{p=0}^{T}\left\|\hat{y}_{k_{i} \rightarrow k_{i}+p}-y_{k_{i}+p}\right\|^{2}_2  
\end{split}
\end{align}
with $B \subset\{1,2, \ldots, N\}$. This reformulation offers the possibility of using powerful optimization algorithms such as Adam \cite{adam}.

\vspace{.1cm}
\section{Experiments and results}\label{section_experiments}
\vspace{.1cm}
We demonstrate the performance of the proposed method on the simulation example of an autonomous Van der Pol oscillator with full state measurements and the Silverbox benchmark system, which is a real-world setup with only input-output data available.
\subsection{Van der Pol}
\vspace{.1cm}
We consider the dynamics of an unforced Van der Pol oscillator \cite{khalil_nl}:
\begin{align}
    \begin{split}
        \dot{x}_1(t) &= x_2(t)\\
        \dot{x}_2(t) &= \mu(1-x^2_1(t))x_2(t)-x_1(t)
    \end{split}
\end{align}
with $\mu=1$. The continuous time system is discretized using the Runge-Kutta 4 numerical formula with a sampling frequency of $20$ Hz. Training, validation and test trajectories are generated starting from initial conditions that are uniformly distributed $x_0\sim\mathcal{U}(-2,2)$, each trajectory having a length of 501 data points. White Gaussian noise $v$ is added to the simulated state trajectories such that a Signal-to-Noise Ratio (SNR) of 20 dB is achieved per each individual channel (note that the test data is noiseless). For training, 80 sets of trajectories, for validation, 20 sets and, for testing, 10 sets are generated.
\par The lifting function $e_\theta$ given by \eqref{eq:zk_fullstate} (without the input term) is implemented as a feedforward neural network, with 1 hidden layer, 100 nodes and tanh activation.  The parameters are initialized by sampling from a uniform distribution $\mathcal{U}(-\sqrt{m},\sqrt{m})$, with $m=1/\sqrt{n_\text{in}}$ , and $n_{\text{in}}$ represents the number of inputs to the layer. We consider a lifting dimension $n_z=100$, a prediction horizon value $T=149$ and a batch size  of 256. For training, Adam batch optimization \cite{adam} is used, with a learning rate of $\alpha=10^{-4}$ and the exponential decay rates set to: $\beta_1 = 0.7$ and $\beta_2 = 0.9$. We utilize early stopping, as described in \cite{Gerben_l4dc}, by computing the simulation error on the validation data set after each epoch. We then select the parameters for which the validation cost is minimal. After the training phase, the model along the epochs with the lowest simulation error is used for analysis.
\par
\begin{figure}
    \centering
    \includegraphics[width=0.465\textwidth]{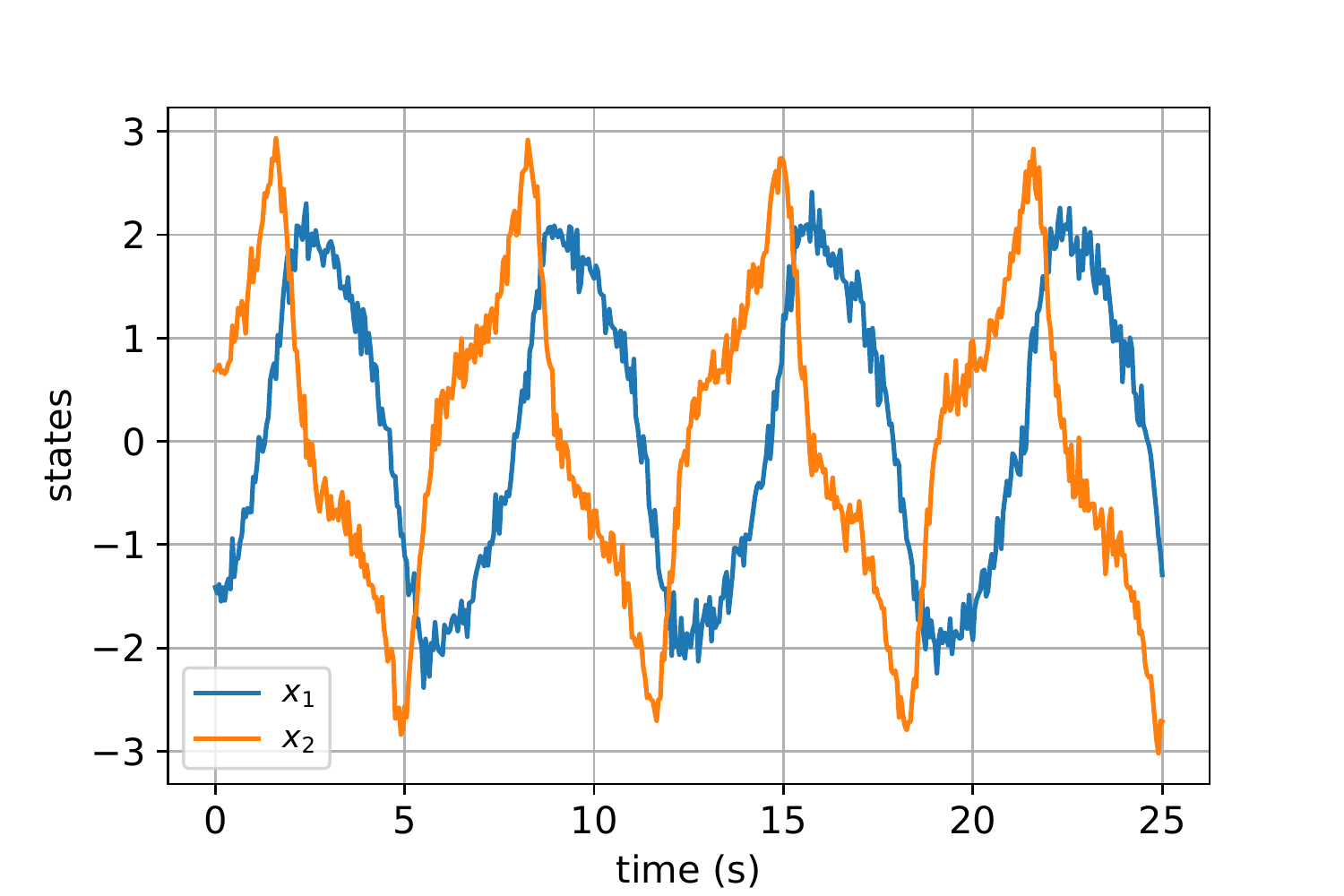}
    \caption{State trajectories of the Van der Pol oscillator with added noise used for training. \vspace{-.5cm}}
    \label{fig:vdp_noisy}
\end{figure}

\begin{figure}
    \centering
    \includegraphics[width=0.465\textwidth]{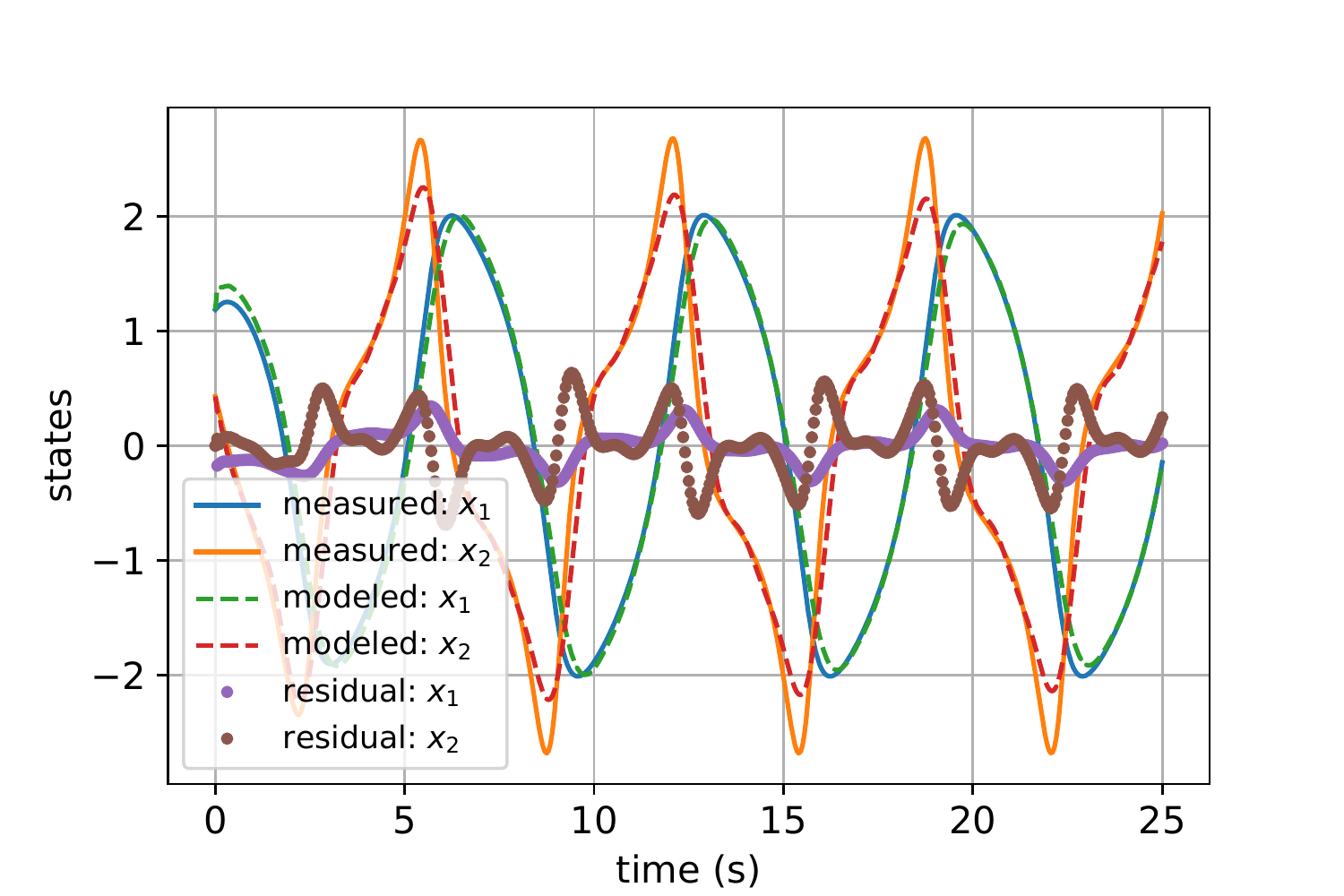}
    \caption{Comparison of the noiseless test data from the Van der Pol oscillator and the the simulation response of the estimated Koopman model. }
    \label{fig:vdp_noiseless}
\end{figure}

\begin{figure}[ht]
    \centering
    \includegraphics[width=0.465\textwidth]{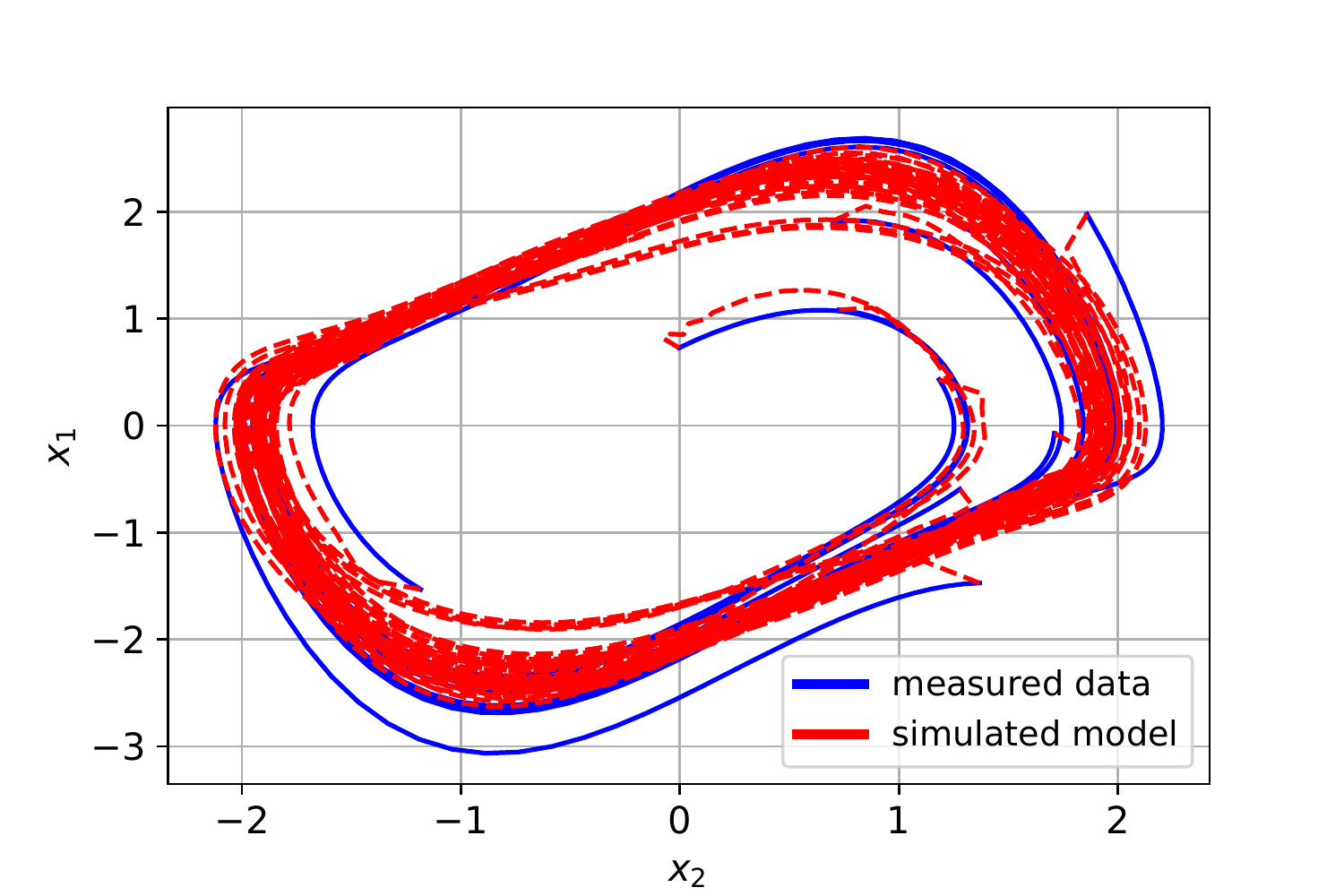}
    \caption{Phase portrait of the noiseless state response of the Van der Pol oscillator in the test data and simulated state trajectories of the identified Koopman model. \vspace{-.15cm}}
    \label{fig:phase_noiseless}
\end{figure}

 Fig. \ref{fig:vdp_noisy} shows a set of noisy time-domain trajectories used as training data. Fig. \ref{fig:vdp_noiseless} depicts one realization of the noiseless test set and the simulated state trajectories of the identified Koopman model, alongside the residuals.  We observe that the proposed Koopman-based encoder is able to capture the oscillating dynamics of the test system with acceptable simulation error. This behavior can also be observed in the phase portrait depicted in Fig. \ref{fig:phase_noiseless}. The quality of the model is assessed in terms of the Normalized Root Mean Square (NRMS) and RMS errors:
\begin{equation}
\mathrm{NRMS}=\frac{\mathrm{RMS}}{\sigma_{y}}=\frac{\text{mean}\left(\sqrt{\frac{1}{M-k_0+1} \sum_{k=k_{0}}^{M}\left\|\hat{y}_{k}-y_{k}\right\|_{2}^{2}}\right)}{\sigma_{y}}
\end{equation}
where the total RMS error is computed as the mean of the RMS error per section of data and  $\sigma_y$ is the standard deviation of all test outputs. $M$ is the total length of a section of test data and $k_0$ is the starting point ($k_0 = \max(n_a,n_b)+1$) for the input-output case and $k_0=1$ for the full state availability case). In terms of this error measure, the following results are obtained on the test data by the estimated Koopman model:
\begin{equation*}
   \mathrm{NRMS}=0.12\qquad\mathrm{RMS}=0.18,
\end{equation*}
where the given error measures are the mean NRMS and RMS over the two state trajectories. These errors can be mostly attributed to the mismatch at the peaks of the sharp rising slopes, as seen in Fig. \ref{fig:vdp_noisy}. However, the identified linear system based on the full state Koopman encoder is able to represent the nonlinear dynamics of the Van der Pol oscillator, successfully recovering the limit cycle. The results are quite satisfactory given that noisy training and validation data sets with $10\%$ noise (in terms of power) are used.
\subsection{Siverbox}
\begin{figure}
    \centering
    \hbox{\hspace{+1em}\includegraphics[scale=0.51]{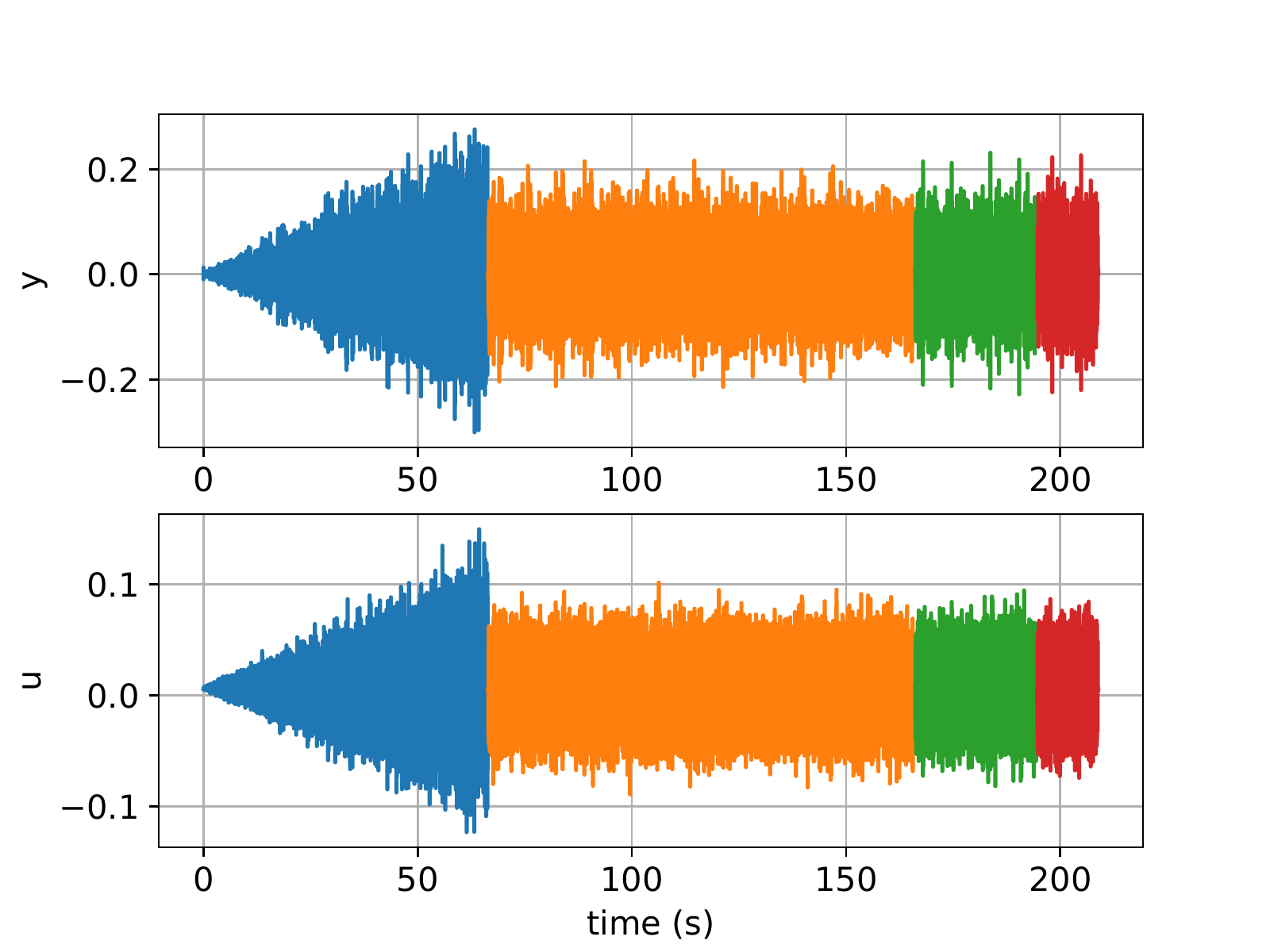}}
    \caption{Separation of the Silverbox data: arrowhead test \protect\blueline, train \protect\orangeline, validation \protect\greenline, test \protect\redline .}
    \label{fig:data_silverbox}
\end{figure}

\vspace{.1cm}
To illustrate the capabilities of the proposed Koopman encoder structure when no state measurements are available, we use measured data from the Silverbox benchmark \cite{Silverbox}, which is a real-world electrical implementation of a mass-spring-damper system with a cubic spring, similar to the forced Duffing oscillator. The first part of the input signal is a filtered Gaussian noise sequence with linearly increasing amplitude, and the rest is generated as a multisine signal. Fig. \ref{fig:data_silverbox} shows the separation of data, with the note that both the multisine and filtered Gaussian (arrowhead part) are used for testing and assessing the quality of the identified model. 
\par For this experiment, both encoder $e_\theta$ \eqref{eq:zk_io} and nonlinear input function $B(z)$ are implemented through feedforward neural networks, the former with 2 hidden layers and the latter with 1, each having 40 neurons per layer. The encoder settings are: $n_z=20$, $T = 49$, $n_a=n_b=10$ and a batch size of 256 is used. The initial parameters are sampled from a uniform distribution as detailed in the  Van der Pol example and the same Adam batch optimization algorithm is used. The learning rate is set to $\alpha=10^{-3}$ and the exponential decay rates are chosen as $\beta_1=0.9$ and $\beta_2=0.999$.
\par
\begin{figure}
    \centering
    \hbox{\hspace{+1em} \includegraphics[scale=0.51]{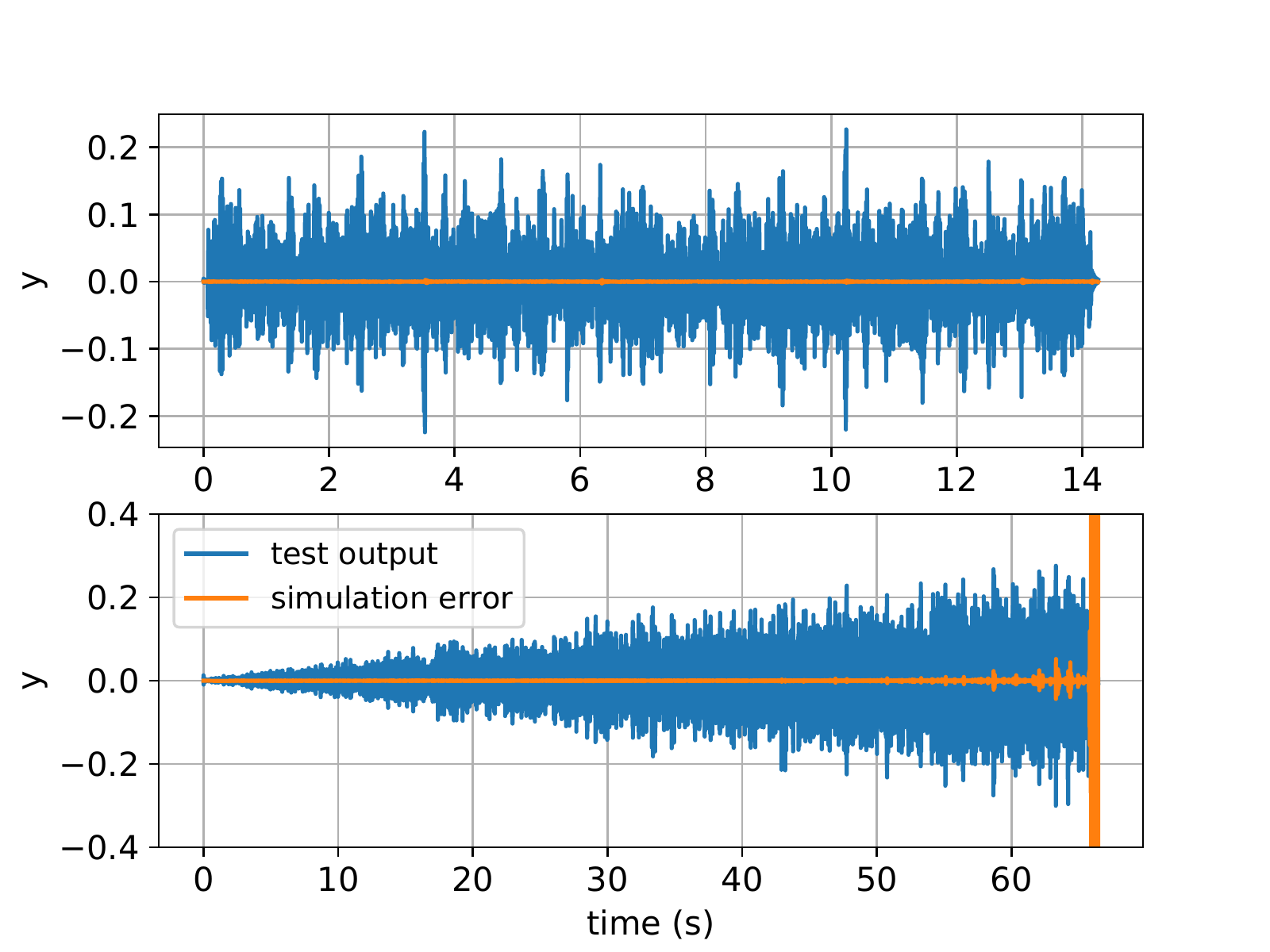}}
    \caption{Silverbox multisine test (top) and arrowhead test (bottom).}
    \label{fig:silverbox-output}
\end{figure}

 As illustrated in Fig. \ref{fig:silverbox-output}, the identified model can accurately represent the dynamics of the original system, when a multisine test signal is applied. However, when the arrowhead test input data is applied, the error highly increases towards the end of the simulation, as Fig. \ref{fig:silverbox-output} depicts. The problem in the extrapolation region is due to a mismatch between the representation of the nonlinearity in the model and the true polynomial nonlinearity. Methods which explicitly use polynomial basis may perform better in this regard \cite{Gerben_l4dc}. Table \ref{table:1} presents the NRMS and RMS error values. If we discard the extrapolation region of the arrowhead test signal, the obtained errors are comparable to the state of the art \cite{Gerben_l4dc}.  \vspace{.3cm}
\vspace{-.5cm}
\begin{table}[!ht]
\vspace{.2cm}
\centering
\caption{Error measures}
\resizebox{0.9\columnwidth}{!}{
\begin{tabular}{|l|c|c|}
\hline
 &  NRMS &  RMS (V)    \\
 \hline\hline
 Test & 0.00552  &  0.00029 \\
 \hline
 Arrowhead &  229.411 &    12.2502 \\
 \hline
 Arrowhead - no extrapol. & 0.00811 & 0.00033\\
 \hline
\end{tabular}
 \vspace{-5cm}
}
\label{table:1}
\end{table}
\vspace{-.5cm}

\vspace{.1cm}
\section{CONCLUSION}\label{section_conclusion}
\vspace{.1cm}
The present paper formulates a Koopman identification method as a nonlinear identification problem, using a neural network estimator consistent with the inverse of the constructability map. Furthermore, the effect of the input is accounted for in the Koopman model through an input-affine description. We have shown that this approach can successfully capture the dynamics of the underlying nonlinear system through motivating examples, for both full and partial state availability. 

\vspace{-.1cm}
\begin{thebibliography}{99}
\vspace{.1cm}
\bibitem{Schoukens_NI}
Schoukens, J. and  Ljung, L., “Nonlinear System Identification: A User-Oriented Roadmap,” \textit{IEEE Control Systems Magazine}, Volume 39, no. 6, pp. 28-99, 2019
\bibitem{Gerben_l4dc}
Beintema, G., Tóth, R. and Schoukens., M., “Nonlinear state-space identification using deep encoder networks,” \textit{Proceedings of Learning for Dynamics and Control (L4DC)}, 2021
\bibitem{Gerben_paper}
Beintema, G., Tóth, R. and Schoukens., M., “Non-linear State-space Model Identification from Video Data using Deep Encoders,” \textit{19th IFAC Symposium on System Identification (SYSID)}, 2021
\bibitem{book_koopman}
Mauroy, A., Mezić, I. and Susuki, Y., \textit{The Koopman Operator in Systems and Control}, Springer International Publishing, 2020
\bibitem{bonnert_nn}
Bonnert, M. and Konigorski, U., "Estimating Koopman Invariant Subspaces of Excited Systems Using Artificial Neural Networks," \textit{IFAC-PapersOnLine}, Volume 53, Issue 2, pp. 1156-1162, 2020
\bibitem{lusch_nn}
Lusch, B., Kutz, J.N. and Brunton, S.L., "Deep learning for universal linear embeddings of nonlinear dynamics," \textit{Nature Communications,} Volume 9, Article no. 4950, 2018
\bibitem{deepkoco_nn}
Heijden, B.V.D., Ferranti, L., Kober, J. and Babuška, R., “DeepKoCo: Efficient latent planning with an invariant Koopman representation,” \textit{arXiv preprint}, arXiv:2011.12690, 2020
\bibitem{Otto}
Otto, S.E. and Rowley, C.W., "Linearly Recurrent Autoencoder Networks for Learning Dynamics", \textit{SIAM Journal on Applied Dynamical Systems}, Volume 18, Issue 1, pp. 558-593, 2019
\bibitem{yeung_nn}
Yeung, E., Kundu, S. and Hodas, N.O., “Learning Deep Neural Network Representations for Koopman Operators of Nonlinear Dynamical Systems,” \textit{American Control Conference (ACC)}, pp. 2832-4839, 2019
\bibitem{Silverbox}
Wigren, T. and Schoukens, J., "Three free data sets for development and benchmarking in nonlinear system identification," \textit{European Control Conference (ECC)}, pp. 2933-2938, 2013
\bibitem{Mauroy_sysid}
Mauroy, A. and Goncalves, J., "Koopman-Based Lifting Techniques for Nonlinear Systems Identification," \textit{IEEE Transactions on Automatic Control}, Volume 65, no. 6, pp. 2550-2565, 2020
\bibitem{rowley_dmd}
Rowley, C., Mezić, I., Bagheri, S., Schlatter, P. and Henningson, D., "Spectral analysis of nonlinear flows," \textit{Journal of Fluid Mechanics}, Volume 641, pp. 115-127, 2009
\bibitem{williams_edmd}
Williams, M.O., Kevrekidis, I.G. and Rowley, C.W., 
"A Data–Driven Approximation of the Koopman Operator: Extending Dynamic Mode Decomposition,"
\textit{Journal of Nonlinear Science}, Volume 25, Issue 6, pp. 1307-1346, 2015
\bibitem{Brunton_2016}
Brunton, S.L., Brunton, B.W., Proctor, J.L. and Kutz, J.N., "Koopman Invariant Subspaces and Finite Linear Representations of Nonlinear Dynamical Systems for Control," \textit{The Public Library of Science ONE}, Volume 11, Issue 2, 2016
\bibitem{brunton_overview}
Brunton, S., Budisic, M., Kaiser, E. and Kutz, J., “Modern Koopman Theory for Dynamical Systems,” \textit{arXiv preprint}, arXiv:2102.12086, 2021
\bibitem{korda_mpc}
Korda, M. and Mezic, I., “Linear predictors for nonlinear dynamical systems: Koopman operator meets model predictive control,” \textit{Automatica}, Volume 93, pp. 149-160, 2018
\bibitem{khalil_nl}
Khalil, H.K., \textit{Nonlinear Systems}, Third Edition, Prentice Hall, 2002
\bibitem{nl_obs_1982}
Nijmeijer, H., "Observability of autonomous discrete time non-linear systems: a geometric approach," \textit{International Journal of Control}, Volume 36, no. 5, pp. 867-874, 1982
\bibitem{adam}
Kingma, D.P. and Ba, J., "Adam: A Method for Stochastic Optimization," \textit{International Conference on Learning Representations (ICLR)}, 2015
\bibitem{takeishi_noise}
Takeishi, N., Kawahara, Y. and Yairi, T., "Learning Koopman Invariant Subspaces for Dynamic Mode Decomposition," \textit{International Conference on Neural Information Processing Systems (NIPS)}, 2017
\bibitem{lpv_book}
Mohammadpour, J. Scherer and C.W., \textit{Control of Linear Parameter Varying Systems with Applications}, Springer-Verlag New York, 2012
\bibitem{surana_obs}
Surana, A., "Koopman Operator Based Observer Synthesis for Control-Affine Nonlinear Systems," \textit{IEEE 55th Conference on Decision and Control (CDC)}, pp.6492-6499, 2016
\end{thebibliography}
\end{document}